\documentclass[a4paper, 12pt]{article}

\usepackage{float}
\usepackage[T1]{fontenc}
\usepackage{lmodern}

\usepackage{setspace}
\usepackage{pdflscape}
\usepackage{longtable}
\usepackage{wasysym}
\usepackage{graphicx}
\usepackage{hyperref}
\usepackage{placeins}
\usepackage{tabularx}
\usepackage{booktabs}
\FloatBarrier
\hypersetup{
    colorlinks=true,
    linkcolor=blue,
    urlcolor=blue,
    citecolor=blue}
\usepackage{float}
\usepackage[
backend=biber,
style=alphabetic,
]{biblatex}
\begin{document}

\baselineskip 12pt

\begin{center}
\textbf{\Large Convergent Representations of Linguistic Constructions in Human and Artificial Neural Systems} \\

\vspace{1.5cc}
{ \sc Pegah Ramezani$^{1,2}$, Thomas Kinfe $^{3,4}$, Andreas Maier $^{2}$, \\ Achim Schilling $^{2,3,4,5}$, Patrick Krauss$^{2,3,4,5}$}\\

\vspace{0.3 cm}

{\small $^{1}$Department of English and American Studies, University Erlangen-Nuremberg, Germany \\ $^{2}$Pattern Recognition Lab, University Erlangen-Nuremberg, Germany\\ $^{3}$Neuromodulation and Neuroprosthetics, University Hospital Mannheim, University Heidelberg, Germany\\$^{4}$BGU Ludwigshafen, Germany\\ $^{5}$Neuroscience Lab, University Hospital Erlangen, Germany}
 \end{center}
\vspace{1.5cc}

\begin{abstract}
Understanding how the brain processes linguistic constructions is a central challenge in cognitive neuroscience and linguistics. Recent computational studies have shown that artificial neural language models spontaneously develop internally differentiated representations of Argument Structure Constructions (ASCs), generating specific predictions about when and how construction-level information should emerge during language processing. The present study tests these predictions in human neural activity using electroencephalography (EEG). Ten native English speakers listened to 200 synthetically generated sentences balanced across four construction types (transitive, ditransitive, caused-motion, and resultative) while neural responses were recorded. Construction-related activity was analyzed using time-frequency analysis, statistical feature extraction, and machine learning classification. The results reveal construction-specific neural signatures that emerge primarily at sentence-final positions, where argument structure information becomes fully disambiguated, and are most prominent in the alpha frequency band. Pairwise classification further demonstrated reliable differentiation between specific construction pairs, particularly ditransitive and resultative constructions, while other pairs exhibited overlapping neural signatures. Importantly, the temporal emergence and similarity structure of these neural effects closely mirror patterns previously observed in recurrent and transformer-based language models, in which constructional representations arise during integrative stages of processing. Together, these findings provide converging evidence that linguistic constructions are neurally encoded as distinct form-meaning mappings, supporting core assumptions of Construction Grammar and suggesting that both biological and artificial learning systems converge on similar representational solutions during predictive language processing. More broadly, this convergence is consistent with the idea that learning systems may discover stable regions within an underlying representational landscape - recently referred to as a Platonic representational space - that constrains which linguistic abstractions emerge as cognitively and computationally efficient forms.

\vspace{0.95cc}
\parbox{24cc}{\it Construction Grammar (CxG); Argument Structure Constructions (ASC); EEG; Neurolinguistics; Large Language Models (LLMs); Recurrent Neural Networks; Neural Language Models; Neural Oscillations; Brain–AI Correspondence}
\end{abstract}
\onehalfspacing

\section{Introduction}

Understanding how the brain processes and represents language is a fundamental challenge in cognitive neuroscience \cite{pulvermuller2002neuroscience}. 
This paper adopts a usage-based constructionist approach, which views language as a system of form-meaning pairs (constructions) that link patterns to specific communicative functions \cite{goldberg2009nature, goldberg2003constructions}. 
Argument Structure Constructions (ASCs), such as transitive, ditransitive, caused-motion, and resultative constructions, are significant for language comprehension and production \cite{goldberg1995constructions, goldberg2006constructions, goldberg2019explain}. 
These constructions are key to syntactic theory and essential for constructing meaning in sentences. 
Exploring the neural and computational mechanisms underlying the processing of these constructions can yield significant insights into language and cognition \cite{pulvermuller2012meaning, pulvermuller2021biological, henningsen2022modelling, pulvermuller2023neurobiological}.

In neurolinguistics, growing evidence suggests that both syntactic and semantic combinatorial processes recruit a left-lateralized fronto-temporal network, including inferior frontal and superior temporal regions \cite{fedorenko2011functional}. Event-related potential (ERP) and time-frequency studies further indicate that different aspects of syntactic integration are reflected in neural oscillations, particularly in the alpha and beta bands \cite{tanner2014erps}. However, despite extensive work on morphosyntax, agreement, and phrase structure, surprisingly little is known about the neural representation of constructions as theorized in Construction Grammar. Unlike traditional syntactic theories, CxG posits that constructions themselves — not only lexemes — are psychologically real units of processing. Direct neural evidence for this claim remains scarce.

Recent advances in computational linguistics have begun to shed light on this issue. Neural network language models, including both recurrent architectures and transformer-based models, have been shown to encode abstract syntactic and constructional information in their internal representations \cite{surendra2023word, krauss2025word}, and to be able to predict language processing in the brain \cite{schrimpf2020artificial}. In two recent studies, we demonstrated that a biologically inspired recurrent neural language model \cite{ramezani2025analysisLSTM} and a transformer model \cite{ramezani2025analysisBERT} spontaneously form distinct clusters for different ASCs, even under tightly controlled syntactic conditions and minimal semantic variation. These findings provide computational support for the idea that constructions emerge as discrete representational units during predictive language processing.

What remains unclear is whether human neural activity mirrors these computational representations. If both artificial and biological neural systems converge on similarly differentiated representations of ASCs, this would provide strong evidence for the cognitive reality of constructions and for shared algorithmic principles underlying language processing in brains and machines.

The present study addresses this gap by investigating whether human EEG signals differentiate between four ASCs — transitive, ditransitive, caused-motion, and resultative — during naturalistic auditory sentence comprehension. Using time-frequency analysis, statistical feature extraction, and machine learning classification, we test whether construction-specific neural signatures can be detected and whether their discriminability parallels the representational structure observed in language models. By directly comparing human electrophysiology with computational models, this work aims to bridge neurolinguistics and artificial intelligence, providing converging evidence for the neural reality of constructions and offering a novel perspective on how form–meaning pairings are processed in the brain.

Recent computational work has demonstrated that artificial neural language models develop internally differentiated representations of Argument Structure Constructions (ASCs) during predictive learning, even in the absence of explicit syntactic supervision \cite{ramezani2025analysisBERT, ramezani2025analysisLSTM}. Across both recurrent and transformer-based architectures, these studies revealed three consistent properties: constructional distinctions emerge primarily at later stages of processing, they depend on the availability of sufficient argument structure information, and they exhibit graded similarity relations reflecting shared event-structural properties between constructions. These findings generate specific predictions for human language processing. If construction-level representations reflect general computational principles rather than architecture-specific mechanisms, comparable patterns should be observable in human neural activity during incremental sentence comprehension. The present study therefore investigates whether EEG signals differentiate between ASC types and whether the temporal emergence, similarity structure, and spectral characteristics of these neural responses align with predictions derived from computational models. By directly linking model-derived hypotheses with electrophysiological data, this work aims to provide converging evidence for constructional representations as a shared organizational principle across artificial and biological language systems.

\section{Methods}

\subsection{Participants}
We recorded the neural responses of 12 participants (8 females, mean age 43 years, range 26-62 years) while they listened to an audiobook. All participants were healthy, right-handed individuals, English native speakers with normal hearing, and reported no history of neurological disorders or substance abuse. The study protocol was approved by the Ethics Committee of the University Hospital Erlangen (No: 22-361-2, PK). All participants provided written informed consent before their inclusion in the study, and all experiments were performed in accordance with relevant guidelines and regulations.

\subsection{Stimulation protocol and EEG recording}
Participants listened to the audio format of sentences generated by GPT-4. This dataset was designed to include sentences that exemplify four distinct ASCs: transitive, ditransitive, caused-motion, and resultative constructions as shown in  Table \ref{tab1}. Each ASC category comprised 50 sentences, yielding a total of 200 sentences.

\begin{table}[h]
\centering
\begin{tabular}{|p{0.20\linewidth}|p{0.28\linewidth}|p{0.28\linewidth}|}
\hline
\textbf{Constructions} & \textbf{Structure} & \textbf{Example} \\
\hline
Transitive & Subject + Verb + Object & The baker baked a cake. \\
\hline
Ditransitive & Subject + Verb + Object1 + Object2 & The teacher gave students homework. \\
\hline
Caused-Motion & Subject + Verb + Object + Path & The cat chased the mouse into the garden. \\
\hline
Resultative & Subject + Verb + Object + State & The chef cut the cake into slices. \\
\hline
\end{tabular}
\caption{Name, structure, and example of each construction}
\label{tab1}
\end{table}

Our materials were restricted to sub-constructions of ASCs so that the sentences varied in construction type without introducing significant structural differences that could obscure compositional effects. The generated text was converted to audio using the Google text-to-speech model\footnote{https://cloud.google.com/text-to-speech}. The total number of sentences was 200, divided into three parts. In the middle of each part, participants could rest as much as they needed.

\begin{figure}[h]
\centering
\includegraphics[width=\linewidth]{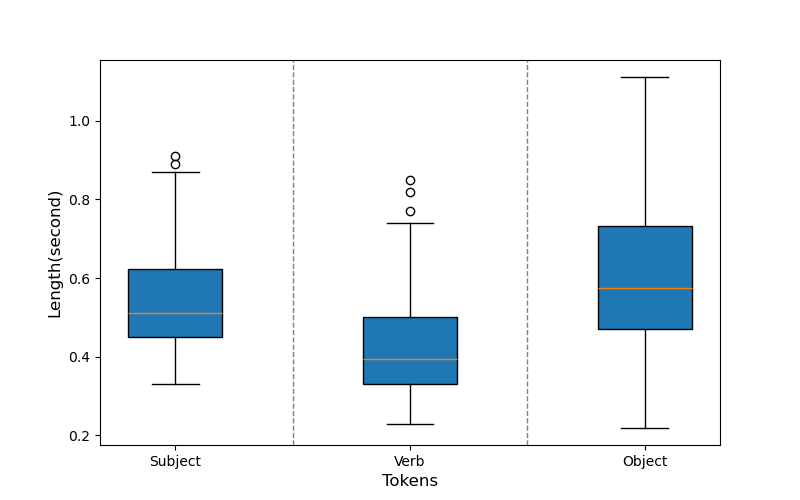}
\caption{Token Duration Variability Across Syntactic Roles. This figure presents boxplots of the durations of subject, verb, and object tokens in the stimulus set. Each boxplot displays the median, interquartile range, whiskers, and outliers, showing that all three syntactic roles exhibit substantial variability in word length. Object tokens show the widest distribution, while verbs are generally shorter. The visualization illustrates the challenge posed by duration differences in natural-language EEG experiments, which motivated the use of maximum-length epoch alignment for ERP analyses and length-independent feature extraction for statistical analyses.}
\label{Token Duration Variability Across Syntactic Roles.}

\end{figure}

A challenge in natural-language EEG experiments is that words within the same syntactic category vary in duration, complicating comparisons of signals of different lengths. As shown in Figure \ref{Token Duration Variability Across Syntactic Roles.}, our stimuli exhibit such length variation. For ERP analyses, we aligned epochs to the maximum duration, whereas for statistical analyses, we used extracted features, which are independent of signal length.

EEG was recorded at 2500 Hz using a Brain Products ActiCAP system with 64 active electrodes arranged according to the international 10–20 system. Electrode impedances were kept below $25 k\Omega$, and an experimenter monitored all recordings.

\subsection{Data preprocessing}

A systematic approach was used to identify and correct faulty channels. Channels showing abnormal characteristics—such as zero variability or unusually high noise—were flagged in the metadata. These channels were then reconstructed using spatial interpolation. In spherical spline interpolation, the scalp is modeled as a smooth surface, and a curvature-minimizing function is fit through the good electrodes; the missing electrode’s value is estimated from this function, ensuring reconstruction without artificial peaks.

Next, an FIR (finite impulse response) band-pass filter between 1–45 Hz was applied using MNE. FIR filters convolve the signal with fixed coefficients so that each output sample is a weighted sum of past inputs, and their linear-phase properties preserve temporal structure. The 1 Hz cutoff removes slow drifts and movement-related artifacts. In comparison, the 45 Hz cutoff attenuates high-frequency noise (e.g., muscle and environmental interference) while retaining theta, alpha, beta, and lower-gamma activity. Filtering before downsampling also prevents aliasing.

The data were downsampled to 250 Hz to reduce computational load. By the Nyquist theorem, sampling must exceed twice the highest frequency present; with a 45 Hz cutoff, anything above 90 Hz is sufficient, so 250 Hz provides a wide margin. Downsampling was performed only after low-pass filtering to ensure no energy above the new 125 Hz Nyquist limit remained.

Independent Component Analysis (ICA; fastICA, 30 components, MNE) was applied to remove ocular artifacts. Following the approach used in Koelbl et al. \cite{koelbl2023adaptive}, components strongly associated with eye movements — most clearly visible in the Fp1 channel — or exceeding an artifact-correlation threshold of 0.2, along with components typically dominated by artifacts (e.g., ICs 0 and 1), were removed. This procedure effectively eliminated blink- and eye-movement–related activity.

After artifact correction, the continuous data were segmented into epochs time-locked to stimulus onset. Baseline correction was applied by subtracting the mean amplitude of the pre-stimulus window from each epoch to compensate for slow drifts. Trials with excessive amplitude fluctuations were automatically rejected.

\subsection{Data analyses}
\begin{figure}[h]
\centering
\includegraphics[width=\linewidth]{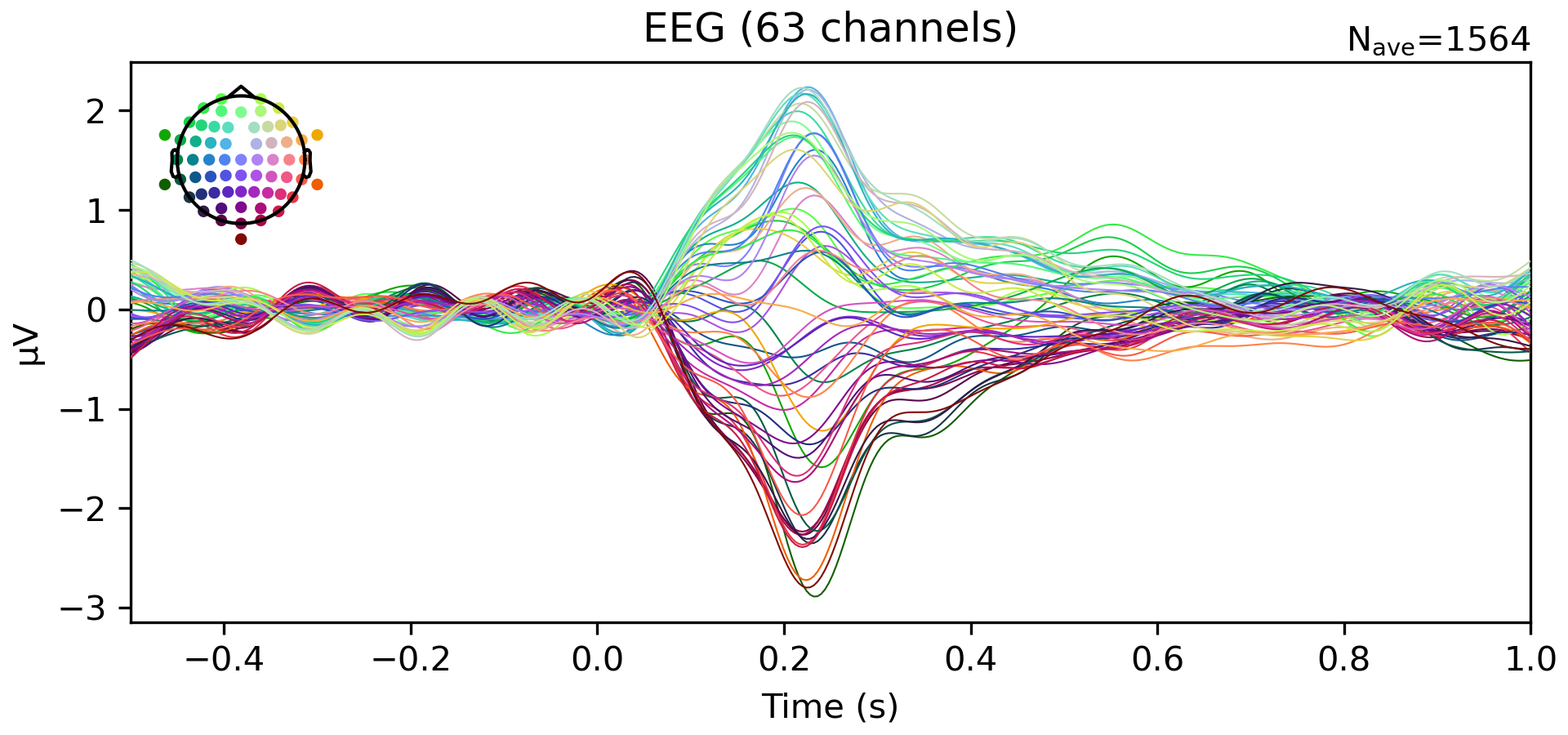}
\caption{ERP Waveforms Time-Locked to Sentence Onset Across All Channels. This figure shows the averaged event-related potentials from 63 EEG channels, aligned to the onset of each sentence. Each colored trace represents one electrode, illustrating the spatial diversity of responses across the scalp. A prominent positive deflection around approximately 200 ms is visible across many channels, corresponding to the P200 component associated with early perceptual and lexical processing. The waveform morphology indicates that the dataset contains robust time-locked neural responses suitable for later time–frequency analyses across construction types.}
\label{ERP Waveforms Time-Locked to Sentence Onset Across All Channels.}
\end{figure}

After preprocessing, we assessed signal quality and verified trigger accuracy by inspecting ERPs time-locked to sentence onset. The averaged ERP across all participants and all four construction types is shown in Figure \ref{ERP Waveforms Time-Locked to Sentence Onset Across All Channels.}, revealing a clear P200 component.

Statistical significance of decoding performance was assessed using permutation testing (1000 label shuffles per participant and construction pair).

For control analyses, cross-validation was additionally performed using verb identity as a grouping factor to prevent lexical overlap between training and test sets.

\FloatBarrier

\section{Results}

\subsection{Time-frequency analysis}

\begin{figure}[h]
\centering
\includegraphics[width=\linewidth]{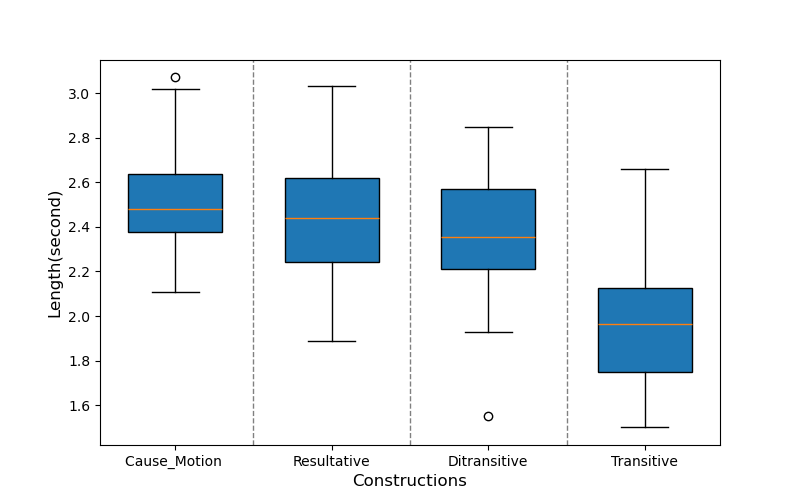}
\caption{Variation in Sentence Length Across Constructions. This figure presents boxplots of sentence durations for the four construction types used in the experiment: Cause\_Motion, Resultative, Ditransitive, and Transitive. Each boxplot shows the median, interquartile range, whiskers, and outliers, illustrating that sentence length varies substantially both within and across constructions. The Transitive sentences tend to be shorter on average, while the other three constructions cluster around longer median durations. This variability motivates standardizing epoch length in EEG analyses, ensuring that neural responses can be aligned and compared despite differences in stimulus duration.}
\label{Variation in Sentence Length Across Constructions.}
\end{figure}

Because sentence lengths vary, as shown in Figure \ref{Variation in Sentence Length Across Constructions.}Epoch duration was standardized using the third-quartile length (covering $74\%$ of sentences) to avoid overlap with subsequent stimuli.

Time–frequency representations were computed separately for each construction type. Because the sentences contained different syntactic items, a joint analysis was not possible; however, comparing the four plots allows us to assess whether corresponding sentence positions exhibit shared or construction-specific activity patterns.

\begin{figure}[h]
\centering
\includegraphics[width=\linewidth]{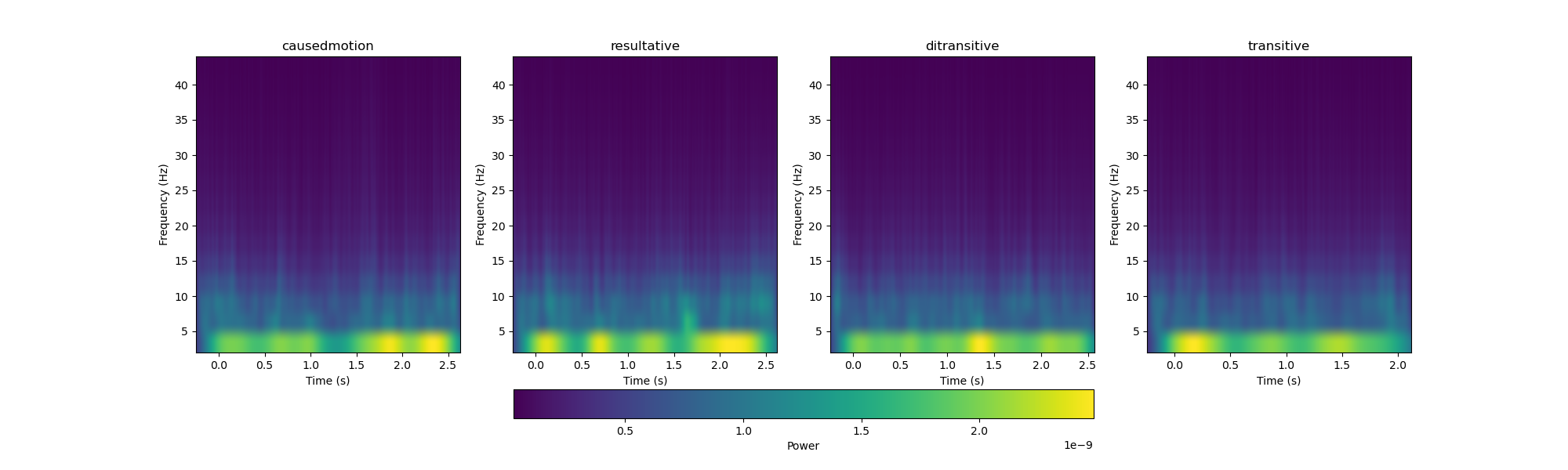}
\caption{Time-frequency representations for each construction type after removing the common average across constructions. The plots show the average Morlet wavelet power (2--45 Hz) for the four construction types (Cause\_Motion, Resultative, Ditransitive, Transitive). Each subplot visualizes power over time and frequency with a shared color scale. Subtracting the common average emphasizes where oscillatory dynamics diverge between constructions, thereby making construction-specific spectral patterns visible across corresponding sentence positions.}
\label{Time–Frequency Representations for Each Construction Type.}
\end{figure}

In Figure \ref{Time–Frequency Representations for Each Construction Type.}It is shown that the strongest construction-specific power occurs at lower frequencies, roughly between 2 and 5 Hz. These values do not reflect overall neural power but rather the residual activity specific to each construction, because the grand average across all constructions has been subtracted. This subtraction removes the shared patterns related to general language comprehension or other concurrent neural processes, leaving only construction-sensitive oscillatory differences. The resulting low-frequency power differences are most pronounced around specific token positions in the sentence.

\FloatBarrier

\subsection{Statistical Analysis}

A statistical analysis was conducted to identify activity patterns associated with each construction by first extracting relevant features from the EEG data. To enable comparison across categories, we focused on epochs aligned to typical syntactic roles—subject, verb, and object. Because sentence durations varied, direct waveform comparison was not possible; instead, we computed multiple statistical features from the broadband signal and from specific frequency bands (delta, theta, alpha, beta, gamma).

Pairwise t-tests were then performed for each construction pair, with p-values corrected using the Benjamini–Hochberg procedure. Features with p < 0.05 were examined to determine which channels, frequency bands, or feature types distinguished the constructions. All statistical tests were conducted separately for each participant.

\begin{figure}[h]
\centering
\includegraphics[width=\linewidth]{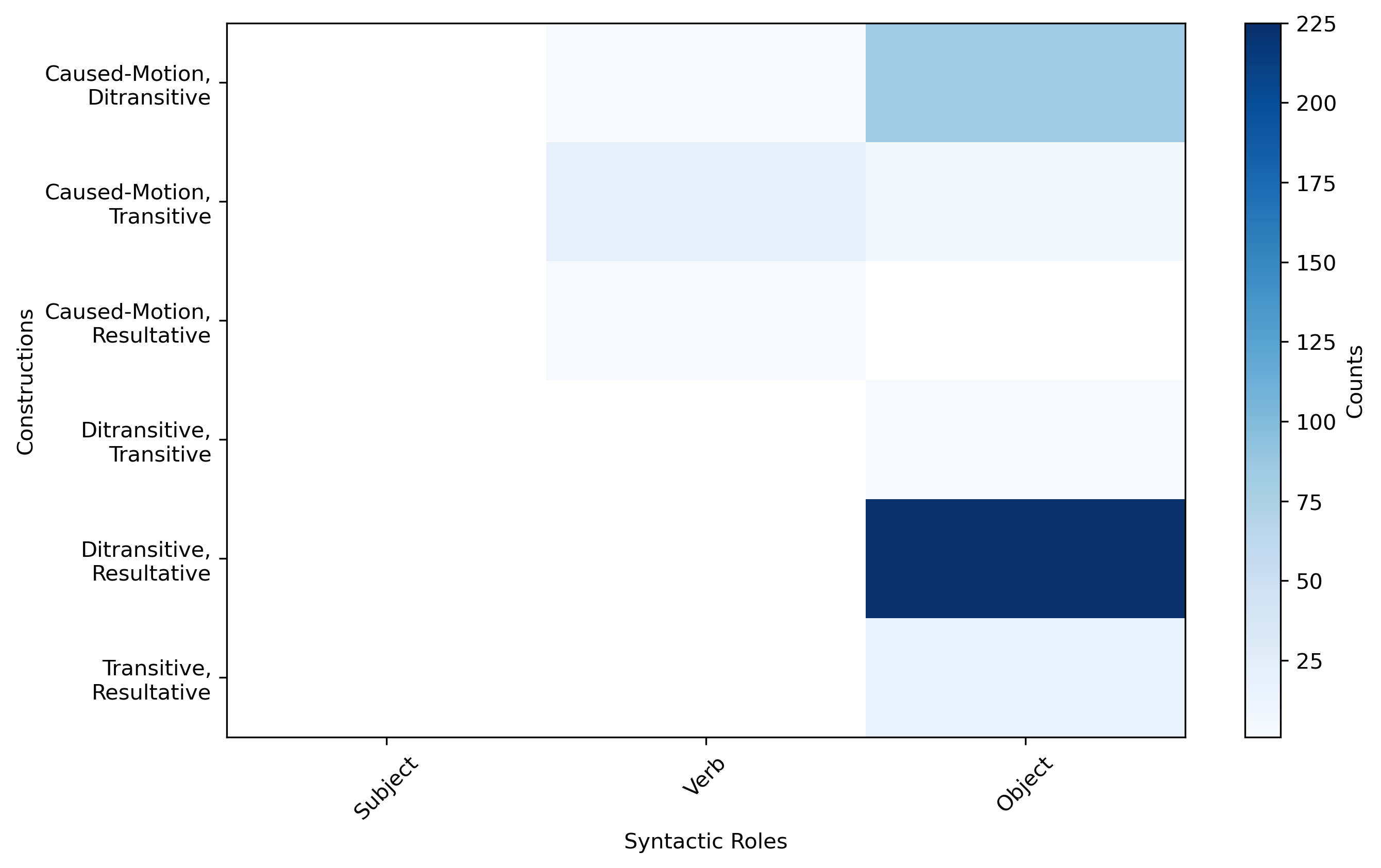}
\caption{Significant EEG Feature Counts Across Construction Pairs and Syntactic Roles. This figure displays a heatmap summarizing the number of EEG features that showed significant differences (p < 0.05, Benjamini–Hochberg-corrected) for each construction pair and each syntactic role: subject, verb, and object. Lighter colors indicate fewer significant features, while darker colors correspond to more distinguishing features. The subject position shows almost no significant differences across all construction pairs, suggesting that early sentence information does not permit construction discrimination. Verb-aligned epochs exhibit only sparse effects, consistent with verbs being less construction-specific. In contrast, object-aligned epochs show substantially more significant features, particularly for the Ditransitive–Resultative pair, indicating that construction-specific neural signatures emerge primarily when listeners have accumulated sufficient contextual information later in the sentence.}
\label{Significant EEG Feature Counts Across Construction Pairs and Syntactic Roles.}
\end{figure}

In Figure \ref{Significant EEG Feature Counts Across Construction Pairs and Syntactic Roles.}, the number of features showing significant differences (p < 0.05) across syntactic roles and construction pairs is summarized.

As expected, no significant differences emerged for the subject position, likely because participants could not distinguish constructions based solely on hearing the subject. A few important differences appeared for verbs, although verbs are generally not construction-specific. In contrast, the object position—occurring near the end of the sentence, when listeners have sufficient contextual information—showed substantially more significant differences than both subject and verb.

Pairwise comparisons revealed that not all construction pairs were distinguishable. The most significant differences appeared between transitive and resultative constructions, followed by caused-motion and ditransitive constructions. Other pairs showed only minor effects, and caused-motion versus resultative yielded no significant differences.

\FloatBarrier
\subsubsection{Statistical Analysis of frequency bands}
Although frequency-band contributions varied across participants, the alpha band was most sensitive to construction differences, followed by beta and delta, with gamma showing the fewest effects. This does not imply that construction processing is confined to these bands; instead, these bands responded most strongly to constructional variation. General oscillatory patterns are evident in the time-frequency analyses.

\begin{figure}[h]
\centering
\includegraphics[width=0.8\linewidth]{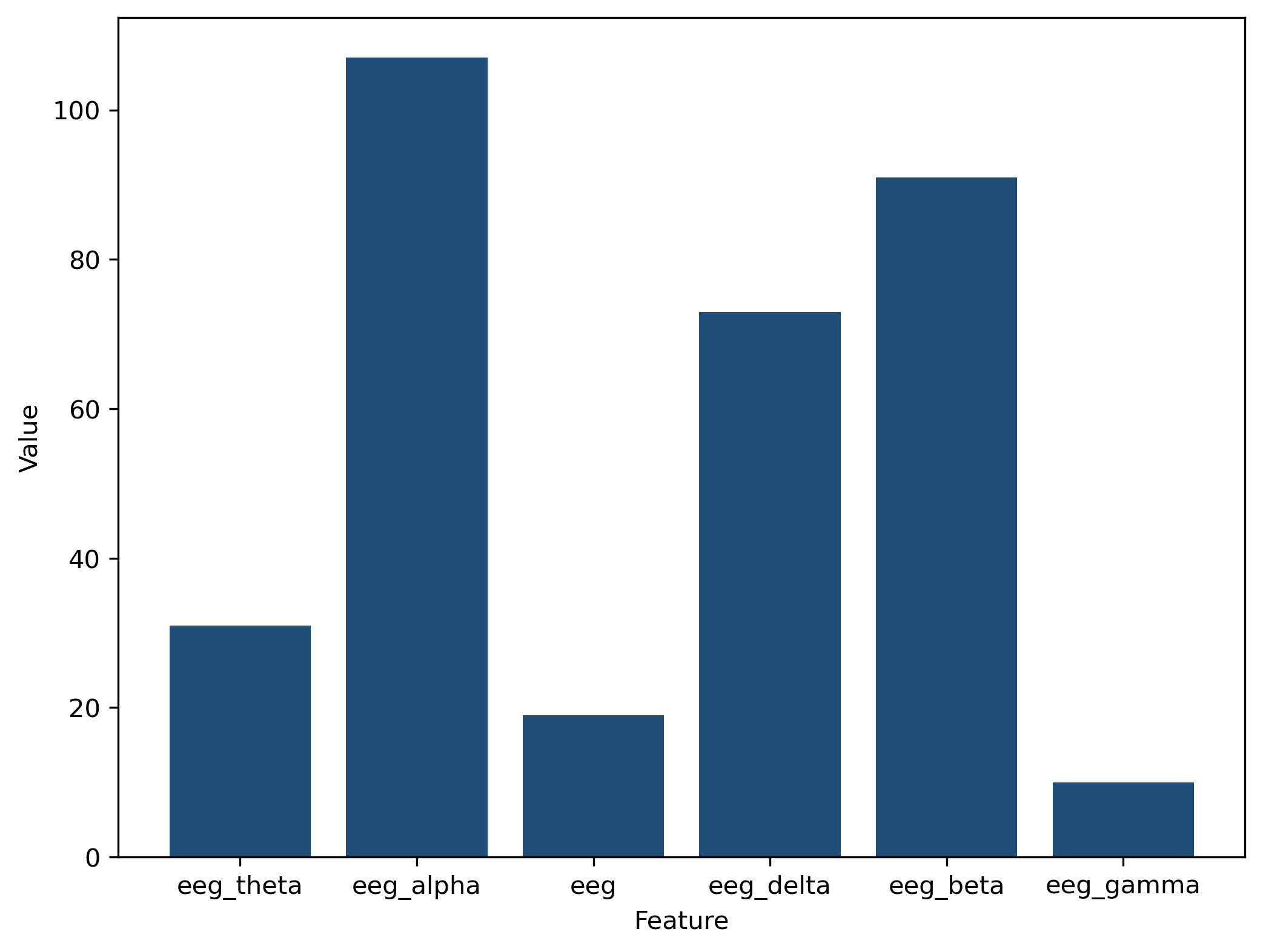}
\caption{Significant EEG Feature Counts by Frequency Band. This figure shows the number of EEG features within each frequency band that exhibited significant differences (p < 0.05, Benjamini–Hochberg corrected) across construction pairs. The alpha band displays the highest number of discriminative features, followed by beta and delta, indicating that these bands were most responsive to constructional variation. Theta and gamma show markedly fewer significant effects. These counts should not be interpreted as exclusive involvement of specific bands in construction processing, but rather as evidence that certain spectral ranges were more sensitive to construction differences. The broader oscillatory patterns underlying these effects are visible in the time–frequency analyses.}
\label{Significant EEG Feature Counts by Frequency Band.}
\end{figure}

\subsubsection{Statistical Analysis of feature type}
The extracted features included mean amplitude, peak amplitude, standard deviation, root-mean-square, kurtosis, and zero-crossing rate. Among these, peak amplitude, signal standard deviation, and kurtosis were most effective in revealing construction-specific differences.

\begin{figure}[h]
\centering
\includegraphics[width=0.8\linewidth]{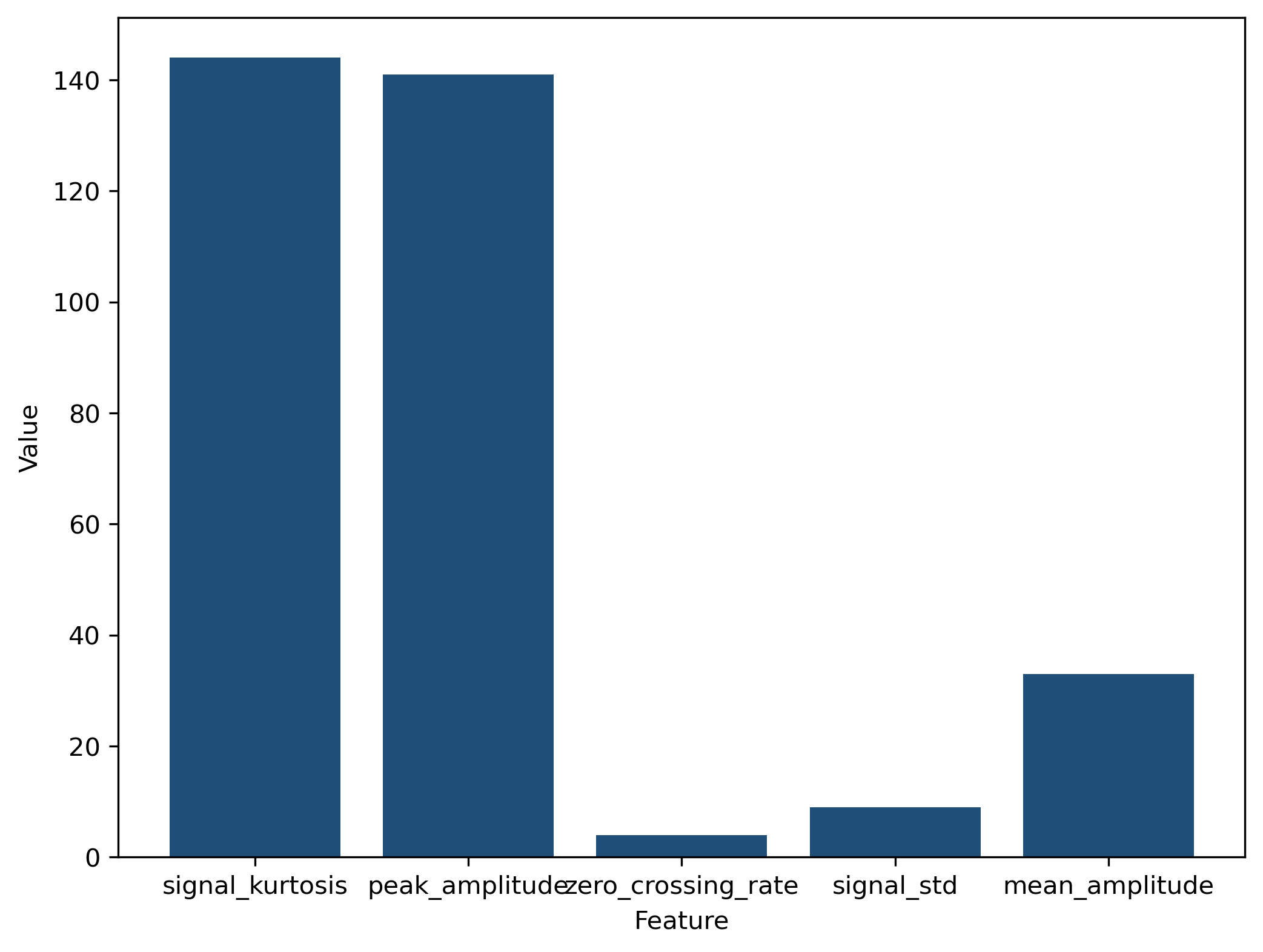}
\caption{Significant Feature Counts for Broadband Signal Measures. This figure shows the number of broadband EEG features that exhibited significant differences (p < 0.05, Benjamini–Hochberg-corrected) across construction pairs for each feature type. Kurtosis and peak amplitude yielded the largest number of distinguishing features, indicating strong sensitivity to construction-related variability. Mean amplitude and standard deviation showed moderate effects, while zero-crossing rate revealed only minimal differences. Together, these results highlight which broadband features were most effective in capturing construction-specific neural patterns.}
\label{Significant Feature Counts for Broadband Signal Measures.}
\end{figure}
\FloatBarrier
\subsection{Assessing Construction Separability via SVM Classification}

For classification, we chose a simple model—not to maximize accuracy, but to test whether construction differences were detectable at all—and used an SVM. We employed a pairwise classification approach, training an SVM with an RBF kernel and evaluating performance using Leave-One-Out (LOO) cross-validation. Accuracy, F1-score, and recall were computed for each construction pair to assess how well the corresponding brain responses could be distinguished.

\subsection{Token classification}

We implemented a classification task to assess whether the model could distinguish between constructions. Because signal lengths varied, we focused on two previously effective features—signal kurtosis and peak amplitude—which were extracted from subject, verb, and object epochs.

\subsubsection{Pairwise decoding results}
When classifying all four constructions simultaneously, the model achieved an average accuracy of $30\%$, only slightly above the $25\%$ chance level. Consistent with our statistical results, not all classes were separable. We therefore switched to pairwise classification to examine each construction pair individually and obtain a clearer picture of their discriminability.

\begin{figure}[h]
\centering
\includegraphics[width=\linewidth]{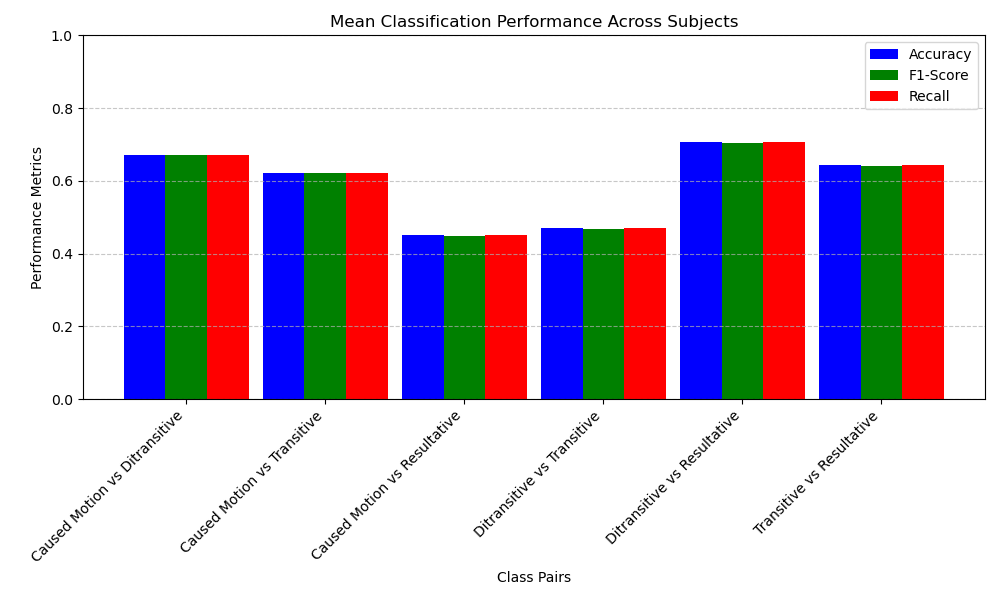}
\caption{Classification Performance for Each Construction Pair. This figure shows the mean accuracy, F1-score, and recall obtained from pairwise SVM classification across subjects for all construction pairs. Each bar group represents one pair of constructions, and the three colors indicate the different evaluation metrics. Pairs such as Ditransitive vs.\ Resultative achieve the highest performance, indicating stronger discriminability in the underlying EEG patterns. In contrast, pairs like Caused-Motion vs.\ Resultative and Ditransitive vs.\ Transitive show lower performance, consistent with the statistical analyses suggesting limited separability. The results demonstrate that while some constructions can be reliably distinguished from one another, others exhibit overlapping neural signatures.}
\label{Classification Performance for Each Construction Pair.}
\end{figure}

In the two-class setting, the chance level is $50\%$, so accuracies at or below this threshold indicate no meaningful separation between classes. In our results, caused-motion vs. resultative and ditransitive vs. transitive show low accuracies, consistent with the statistical analyses. In contrast, the ditransitive vs. resultative pair yields the highest accuracy, further reinforcing the statistical findings.

\subsubsection{Group-level inference}
Classification was performed separately for each participant, and decoding performance was evaluated for each pairwise construction contrast based on object-aligned epochs. Because multi-class classification remained close to chance level, subsequent analyses focused on pairwise contrasts, which provide a more sensitive assessment of construction discriminability.

Decoding performance was summarized at the group level by averaging accuracy across participants. To ensure that group averages reflected reliable above-chance decoding at the participant level, accuracy was expressed relative to chance (accuracy minus chance level), and subject-level inference was performed across participants using Wilcoxon signed-rank tests against zero. This approach allows assessing whether decoding performance was consistently above chance across participants rather than being driven by a subset of individuals. The resulting statistics for each construction pair are reported below.

\begin{table}[ht]
\centering
\caption{Wilcoxon signed-rank test results (accuracy minus chance).}
\label{tab:wilcoxon_acc_minus_chance}
\begin{tabularx}{\textwidth}{l *{5}{>{\centering\arraybackslash}X}}
\toprule
\textbf{Label pair} & \textbf{Mean} & \textbf{Median} & \textbf{W} & \textbf{p} & \textbf{Rank-biserial r} \\
\midrule
Caused Motion vs Ditransitive & 0.1718 & 0.1700 & 66.0 & \textbf{0.000488} & 1.0000 \\
Caused Motion vs Transitive   & 0.1218 & 0.1300 & 66.0 & \textbf{0.000488} & 1.0000 \\
Caused Motion vs Resultative  & -0.0491 & -0.0300 & 17.0 & 0.926270 & -0.4848 \\
Ditransitive vs Transitive    & -0.0309 & 0.0000 & 18.0 & 0.833560 & -0.3455 \\
Ditransitive vs Resultative   & 0.2055 & 0.2000 & 66.0 & \textbf{0.000488} & 1.0000 \\
Transitive vs Resultative     & 0.1427 & 0.1400 & 66.0 & \textbf{0.000488} & 1.0000 \\
\bottomrule
\end{tabularx}
\end{table}

\subsubsection{Permutation-based validation}

To further assess whether decoding performance exceeded chance levels, statistical significance was additionally evaluated using permutation testing. For each participant and construction pair, class labels were randomly shuffled 1000 times, and decoding accuracy was recomputed to generate a contrast-specific null distribution. Observed decoding accuracies were then compared against these null distributions to estimate empirical permutation-based significance.

Across construction pairs, observed accuracies consistently exceeded the corresponding permutation-based null accuracies for contrasts that showed strong decoding performance, whereas contrasts with low decoding accuracy overlapped with the null distribution. These results confirm that the observed discriminability cannot be explained by chance-level fluctuations in the data. Summary statistics across construction pairs are reported in Table \ref{tab:avg_by_label_pair}.

\begin{table}[ht]
\centering
\caption{Grand average decoding results across subjects ($n=11$) for each pairwise construction contrast. Reported are the mean observed leave-one-out accuracy, the mean permutation null accuracy (1000 label shuffles), and the mean empirical permutation $p$-value (averaged across subjects).}
\label{tab:avg_by_label_pair}
\begin{tabularx}{\textwidth}{l *{3}{>{\centering\arraybackslash}X}}
\toprule
\textbf{Label pair} & \textbf{Observed acc.} & \textbf{Perm. acc.} & \textbf{Perm. $p$} \\
\midrule
Caused Motion vs Ditransitive & 0.673 & 0.454 & \textbf{0.0136} \\
Caused Motion vs Transitive   & 0.625 & 0.455 & \textbf{0.0386} \\
Caused Motion vs Resultative  & 0.454 & 0.451 & 0.4980 \\
Ditransitive vs Transitive    & 0.471 & 0.456 & 0.4565 \\
Ditransitive vs Resultative   & 0.705 & 0.456 & \textbf{0.0037} \\
Transitive vs Resultative     & 0.642 & 0.455 & \textbf{0.0181} \\
\bottomrule
\end{tabularx}
\end{table}

\begin{figure}[H]
\includegraphics[width=\linewidth]{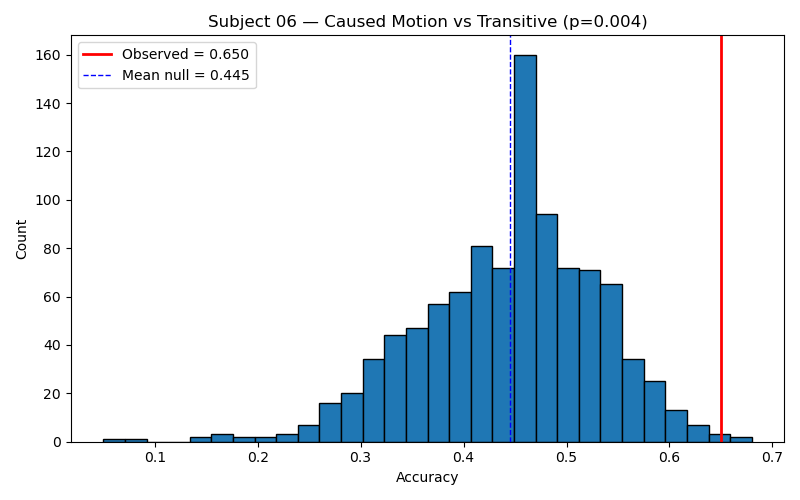}
\caption{Permutation test null distribution for Subject 06 (Caused Motion vs Transitive). The histogram shows decoding accuracies obtained from 1000 label permutations (null distribution). The solid red line marks the observed accuracy (0.650), and the dashed blue line marks the mean null accuracy (0.445). The permutation p value (p = 0.004) is the proportion of permuted accuracies greater than or equal to the observed accuracy.}
\label{permutesbj4}
\end{figure}

\subsubsection{Control analysis: verb leakage}

To exclude the possibility that decoding performance was driven by lexical overlap rather than construction-level information, we performed an additional control analysis addressing potential verb leakage between training and test data. Inspection of the stimulus set revealed that several verbs occurred across multiple sentences and constructions, raising the possibility that classifiers could exploit verb-specific cues rather than constructional structure (see Supplementary Figure X for lexical distribution).

To control for this effect, decoding was repeated using a verb-controlled cross-validation scheme in which all sentences containing the same verb were assigned exclusively to either the training or the test set within each fold. This procedure prevents the classifier from relying on verb-specific information and requires generalization across verbs. Decoding performance under this constraint showed a slight reduction in overall accuracy but preserved the relative pattern of discriminability across construction pairs. In particular, contrasts that exhibited strong separability in the standard analysis remained above chance level.

\begin{figure}[H]
\includegraphics[width=\linewidth]{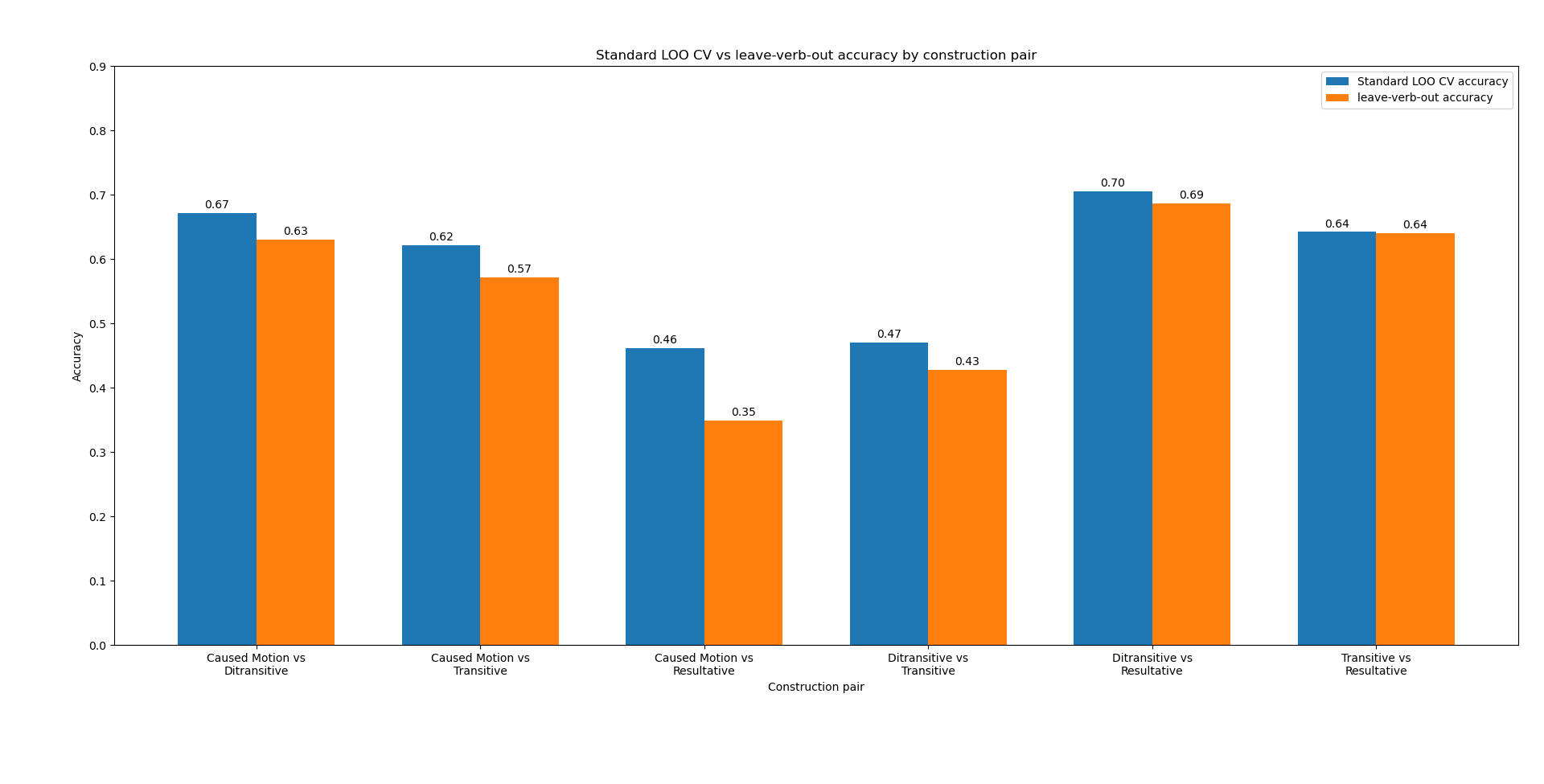}
\caption{Mean binary classification performance across subjects for each pair of constructions for two senarios, one with standard CV and the other one with Leave verb out..}
\label{verb_lexical}
\end{figure}

These results indicate that construction-specific decoding cannot be explained by lexical overlap between constructions and instead reflects construction-level neural differentiation.

\FloatBarrier
\section{Discussion}

\subsection{Model-Derived Predictions and Neural Confirmation}

The present study was designed to test whether representational principles previously identified in computational language models generalize to human neural processing during sentence comprehension. In two preceding studies \cite{ramezani2025analysisLSTM, ramezani2025analysisBERT}, we showed that both a recurrent neural language model and a transformer-based language model develop internally differentiated representations of Argument Structure Constructions (ASCs) despite substantial architectural differences and minimal explicit supervision. Importantly, these computational results do not merely demonstrate that artificial systems can represent constructions; they generate concrete predictions about when and how construction-level information should emerge during incremental language processing. The EEG results presented here allow these predictions to be evaluated at the level of human neural dynamics.

A first prediction concerns the temporal locus of constructional differentiation. In both the recurrent and transformer-based models, construction-specific representations emerged only after sufficient argument structure information had accumulated to determine the underlying event structure of the sentence. Early elements such as subjects or determiners carried little construction-specific information, whereas later elements—particularly those associated with the object or completion of argument structure—showed the strongest differentiation. The EEG findings closely follow this predicted pattern. Construction-specific neural differences were largely absent at early sentence positions and became most pronounced at the object position, where listeners can reliably infer the constructional meaning of the sentence. This convergence suggests that construction processing in the human brain reflects an integrative stage of interpretation rather than early syntactic categorization, supporting usage-based accounts in which constructions emerge from accumulated relational information during comprehension.

A second prediction concerns the internal similarity structure among constructions. Computational analyses revealed that ASC representations are not uniformly separated but instead organized according to graded similarities reflecting shared structural and semantic properties. Some construction pairs formed clearly distinct clusters, whereas others remained partially overlapping. The EEG results exhibit the same qualitative structure: ditransitive and resultative constructions showed robust separability, whereas caused-motion and resultative constructions exhibited substantial overlap. This correspondence indicates that the convergence between artificial and biological systems extends beyond the detection of constructional differences per se and includes the relative geometry of these representations. Such alignment is consistent with the interpretation that both systems converge on similar representational solutions because they are constrained by common demands of event-structure integration during predictive language processing.

A third point of convergence concerns the level of processing at which constructional distinctions become neurally expressed. In computational models, construction-specific information emerges in intermediate and late representational stages associated with contextual integration and abstraction. In the EEG data, construction differences were most prominent in lower-frequency oscillatory activity, particularly in the alpha band, which has repeatedly been associated with integrative and unification processes in language comprehension. The parallel emergence of construction sensitivity at integrative processing stages across systems suggests that ASC representations reflect the construction of higher-level relational meaning rather than the processing of isolated lexical or syntactic features.

Taken together, the results provide converging evidence that construction-level representations follow similar computational principles across artificial neural networks and the human brain. Across recurrent models, transformer architectures, and human electrophysiology, construction distinctions emerge late in processing, depend on the availability of argument structure information, and exhibit comparable similarity relations between construction types. Rather than reflecting architecture-specific mechanisms, this convergence supports the view that constructions constitute stable representational solutions that arise naturally in systems optimized for predictive language processing. In this sense, the present findings strengthen the interpretation that constructional knowledge represents a psychologically real level of linguistic organization, grounded in general principles of learning, prediction, and event-structure integration.

\subsection{Convergence Between Human EEG and Neural Language Models}
The present study demonstrates that human EEG signals differentiate between four Argument Structure Constructions (ASCs)—transitive, ditransitive, caused-motion, and resultative—during naturalistic sentence comprehension. Across complementary analytical approaches, including time–frequency analysis, statistical feature extraction, and machine-learning classification, the results consistently indicate that constructional differences are reflected in neural dynamics. These findings support the interpretation that the human brain represents ASCs as distinguishable form–meaning mappings rather than as purely lexical or syntactic combinations. Importantly, the observed neural patterns do not arise in isolation but align closely with representational structures previously identified in computational models of language processing, including both recurrent neural networks and transformer-based architectures. Taken together, this convergence strengthens the constructionist assumption that ASCs constitute psychologically real units of linguistic knowledge and suggests that similar abstraction principles may operate across biological and artificial neural systems.

A central contribution of the present work lies in the correspondence between neural signatures observed in EEG and the internal representational organization reported in computational models. In both the LSTM and BERT studies, sentence representations formed distinct clusters according to construction type, even under tightly controlled syntactic conditions and minimal semantic variability. The EEG findings exhibit a comparable structure of discriminability, indicating that construction-level distinctions emerge at analogous stages of processing across systems.

First, constructional differentiation was strongest at sentence-final positions, particularly at the object role. This mirrors the computational findings, where construction-specific information was most pronounced in representations associated with later integrative stages of processing. In recurrent models, hidden-state representations corresponding to the object position showed maximal separability between constructions, while in the transformer model object-related embeddings provided the most informative signal for construction identification. The EEG results show the same temporal profile: construction-specific neural differences emerge most clearly when sufficient contextual information has accumulated to determine the underlying event structure. The stability of this effect across permutation testing and verb-controlled decoding further indicates that the observed differentiation reflects construction-level processing rather than lexical confounds or statistical fluctuations.

Second, the similarity relations between constructions follow comparable patterns in EEG and computational models. Not all construction pairs were equally distinguishable; instead, separability reflected structural and semantic relatedness between constructions. Ditransitive and resultative constructions showed the strongest differentiation, whereas caused-motion and resultative constructions exhibited substantial overlap. The presence of this shared similarity structure suggests that convergence occurs not only at the level of detecting constructional differences but also in the organization of representational space itself. Such correspondence is consistent with the interpretation that both biological and artificial systems are constrained by similar demands of event-structure representation during predictive language comprehension.

Third, the spectral characteristics of the EEG effects provide an additional point of alignment with computational findings. Construction-specific differences were most prominent in the alpha frequency band, with additional contributions from beta and delta activity. Alpha oscillations have repeatedly been associated with integrative and unification processes during language comprehension, reflecting the combination of distributed information into coherent meaning representations. This interpretation aligns with the modeling results, in which construction-level distinctions emerged in representational stages associated with contextual integration rather than early lexical encoding. The sensitivity of alpha-band activity to constructional variation may therefore reflect a shared computational principle across systems: the integration of role-based information into higher-level relational meaning.

Taken together, these converging observations suggest that artificial neural networks—despite their architectural differences from biological systems—arrive at internal encodings that resemble those observed in human neural processing. Rather than implying direct mechanistic equivalence, this convergence indicates that both classes of systems may discover similar representational solutions when optimized for predictive language processing. In this sense, the parallel findings support the view that contemporary language models capture aspects of the computational structure underlying human sentence comprehension and provide a useful framework for generating testable hypotheses about neural language processing.

\subsection{Construction-Specific Neural Signatures and Their Linguistic Interpretation}

While the previous section focused on the convergence between neural and computational representations, the EEG results also allow a more direct interpretation of how constructions themselves are neurally encoded during sentence comprehension. The fact that subject and verb positions exhibited almost no construction-specific differences, whereas objects yielded robust effects, indicates that construction processing is incremental and context-dependent. Subjects do not yet constrain construction type, and verbs — although semantically informative — do not uniquely identify ASC membership across all items. Only when the object (and the associated argument structure) becomes available can listeners infer the construction’s conceptual schema.

The pairwise EEG distinctions further support this view. Transitive and resultative constructions differed strongly, consistent with their divergent event structures. Caused-motion and ditransitive constructions also showed separability, albeit to a lesser degree. However, caused-motion vs. resultative did not differ at all. This is linguistically meaningful: many resultative verbs encode changes of location (e.g., pushed the box open), making them semantically similar to caused-motion constructions. These fine-grained semantic overlaps likely contribute to their neural similarity. This interpretation is further supported by the verb-controlled decoding analysis, in which construction discriminability remained largely preserved when lexical overlap between training and test data was eliminated, indicating that the observed differences cannot be explained by verb-specific cues.

Importantly, these effects emerged primarily at low frequencies (2–5 Hz), which are associated with chunking, phrase-level integration, and hierarchical parsing. This suggests that ASC distinctions arise at the level of conceptual event-structure integration rather than at the phonological or lexical level. Consistent with this interpretation, permutation-based validation confirmed that reliable decoding emerged only for construction pairs showing clear conceptual differences, further suggesting that ASC distinctions reflect higher-level integration processes rather than chance-level fluctuations or surface-level lexical variation

\subsection{Frequency Band Contributions and Time Windows}

From a neurocognitive perspective, the spectral profile of the observed effects provides additional insight into the processing stage at which constructional distinctions arise. Consistent with previous ERP and time–frequency studies on combinatorial syntax and semantics \cite{bastiaansen2006oscillatory}, the alpha band was the most sensitive spectral range for detecting ASC differences. Alpha oscillations have been linked to hierarchical processing, anticipatory unification, and controlled retrieval of structured linguistic information. The additional contributions from beta and delta bands align with their known roles in syntactic prediction and temporal integration.

The timing of these effects - emerging near the object position - further supports the interpretation that ASCs are processed as event-level templates rather than as lexical fragments. The EEG therefore appears to reflect not merely the recognition of individual words but the incremental construction of a relational meaning pattern. This interpretation is further supported by the verb-controlled decoding analysis, which indicates that construction-specific effects persist when lexical overlap between constructions is removed.

\subsection{Individual Differences and Cognitive Variability}

The analyses revealed noteworthy individual variation in the degree and location of construction-specific neural effects. Such variability is consistent with previous ERP research demonstrating differences in morphosyntactic processing strategies among native speakers \cite{tanner2014erps}. Variation in linguistic performance, cognitive control, or attentional strategies can also modulate construction processing \cite{FIORENTINO201879}. This raises a broader theoretical question discussed in Construction Grammar: Do all native speakers share the same inventory of constructions, or are there idiosyncratic, experience-based variants?

Our data support the latter possibility. While all participants showed construction sensitivity at the group level, as confirmed by subject-level inference across decoding analyses, the precise channels, frequency ranges, and feature types that distinguished constructions differed across individuals. This suggests multiple neurocognitive pathways for successful construction comprehension, aligning with usage-based theories that emphasize learning histories and experience-shaped representations. In this sense, the EEG results suggest that constructional representations emerge at the level of event-structure integration, providing a linguistic interpretation of the representational convergence discussed above.

\subsection{Limitations}

As with many EEG studies employing controlled linguistic materials, several methodological limitations should be acknowledged when interpreting the present findings.

First, although the stimulus set was carefully designed, we were unable to control sentence length and lexical content with full precision. This variability introduces noise into the EEG signal and complicates the alignment of neural responses across stimuli.

Second, the dataset was small both in terms of the number of participants and the number of trials per construction type. Such constraints can increase the influence of individual differences and may reduce statistical power, potentially obscuring more subtle construction-specific effects. To mitigate these limitations, decoding results were validated using permutation testing, subject-level statistical inference, and verb-controlled cross-validation, which together reduce the likelihood that the reported effects reflect chance fluctuations or lexical confounds.

\subsection{Methodological Considerations: Sub-Constructions and Experimental Constraints}

The present experimental design deliberately focused on sub-types of ASCs rather than highly diverse surface forms in order to maximize experimental control. This design had clear advantages: it allowed aligning syntactic roles across sentences and isolating constructional effects from extraneous linguistic variability. However, the similarity among sub-constructions—particularly between caused-motion and change-of-state/resultative verbs—also makes strong neural dissociations more difficult to detect.

The fact that we nonetheless observed clear construction-specific signatures, supported by permutation-based validation and verb-controlled decoding analyses, indicates that ASCs have a robust neural footprint even under conservative stimulus control. Future studies may test broader ASC families, examine cross-linguistic generalization, or manipulate semantic dimensions within constructions (e.g., location-changing vs. property-changing resultatives) to further refine the representational landscape.

\section{Conclusion and Outlook}

Our results provide converging evidence that EEG signals carry construction-specific information during real-time sentence comprehension and that the structure of these neural representations closely mirrors those observed in computational language models. This neural–computational correspondence strengthens the case for the cognitive reality of constructions and provides evidence that both brains and artificial neural networks rely on comparable learning principles—incremental prediction, role-based integration, and event-structure abstraction—when processing linguistic input.

Future studies may extend the present work by increasing sample size and broadening the set of constructions examined, thereby supporting more robust generalization and finer distinctions among ASC types. Examining L2 learners represents another promising direction, with the potential to illuminate how construction-specific neural signatures vary across proficiency and learning histories. Additionally, cross-linguistic investigations employing parallel constructions across different native languages could help determine whether ASCs rest on universal cognitive principles or emerge from language-specific experience.

Although our study focused on neural signals, the results have broader implications for behavioral theories of sentence processing. Behavioral studies often show graded effects of argument structure predictability, thematic role assignment, and event-structure inference. The present EEG data suggest that these behavioral patterns may emerge from underlying construction-sensitive neural dynamics that accumulate evidence across the sentence until a constructional interpretation becomes available.

The convergence of three lines of evidence—behavioral patterns reported in psycholinguistics, neural signatures shown here, and emergent representations in recurrent and transformer-based language models—is consistent with the view that ASCs constitute a psychologically and computationally real level of linguistic organization, guiding human sentence comprehension and emerging naturally in predictive learning systems.

More broadly, this convergence suggests that learning systems operating under similar computational constraints may converge toward common regions of representational space that function as attractors for useful and stable abstractions. Such convergence has been interpreted as consistent with the idea of Platonic representational regularities, in which certain cognitive structures are discoverable rather than arbitrarily constructed \cite{huh2024platonic, metzner2025core, metzner2025platonic, levin2025platonic, sole2026cognition}. On this view, a Platonic representational space would denote a structured landscape of cognitive and morphological patterns that constrains which forms and meanings are easy to learn, stable across learning systems, and efficient for communication. Identifying these hidden regularities may ultimately offer a unifying framework for understanding how meaning emerges across natural and artificial cognitive systems.

\section{Acknowledgments}
This work was funded by the Deutsche Forschungsgemeinschaft (DFG, German Research Foundation): grants KR\,5148/3-1 (project number 510395418), KR\,5148/5-1 (project number 542747151), KR\,5148/10-1 (project number 563909707) and GRK\,2839 (project number 468527017) to PK, and grants SCHI\,1482/3-1 (project number 451810794) and SCHI\,1482/6-1 (project number 563909707) to AS.

\section{Author Contributions}
PK and AS developed the study protocol. AS and PK supervised the study. PR performed the measurements. PR performed analyses. PR and AS acquired funding. PR, AS, AM and TK provided resources. PR, TK, AS, AM and PK wrote the manuscript.


\printbibliography

@article{metzner2025core,
  title={The Core Of The Scientific Method},
  author={Metzner, Claus},
  journal={Authorea Preprints},
  year={2025},
  publisher={Authorea}
}

@article{metzner2025platonic,
  title={Platonic Attractors},
  author={Metzner, Claus},
  journal={Authorea Preprints},
  year={2025},
  publisher={Authorea}
}

@article{sole2026cognition,
  title={Cognition spaces: natural, artificial, and hybrid},
  author={Sol{\'e}, Ricard and Seoane, Luis F and Pla-Mauri, Jordi and Bennett, Michael Timothy and Hochberg, Michael E and Levin, Michael},
  journal={arXiv preprint arXiv:2601.12837},
  year={2026}
}

@article{levin2025platonic,
  title={Platonic space: where cognitive and morphological patterns come from (besides genetics and environment)},
  author={Levin, M},
  journal={Forms of life, forms of mind. March},
  volume={9},
  year={2025}
}

@inproceedings{koelbl2023adaptive,
  title={Adaptive ica for speech eeg artifact removal},
  author={Koelbl, Nikola and Schilling, Achim and Krauss, Patrick},
  booktitle={2023 5th international conference on bio-engineering for smart technologies (biosmart)},
  pages={1--4},
  year={2023},
  organization={IEEE}
}

@incollection{krauss2025word,
  title={Word Class and Syntax Rule Representations Spontaneously Emerge in Recurrent Language Models},
  author={Krauss, Patrick and Surendra, Kishore and Stoewer, Paul and Maier, Andreas and Metzner, Claus and Schilling, Achim},
  booktitle={Recent Advances in Deep Learning Applications},
  pages={106--122},
  year={2025},
  publisher={Chapman and Hall/CRC}
}

@article{ramezani2025analysisBERT,
  title={Analysis of argument structure constructions in the large language model BERT},
  author={Ramezani, Pegah and Schilling, Achim and Krauss, Patrick},
  journal={Frontiers in Artificial Intelligence},
  volume={8},
  pages={1477246},
  year={2025},
  publisher={Frontiers Media SA}
}

@article{ramezani2025analysisLSTM,
  title={Analysis of argument structure constructions in a deep recurrent language model},
  author={Ramezani, Pegah and Schilling, Achim and Krauss, Patrick},
  journal={Frontiers in Computational Neuroscience},
  volume={19},
  pages={1474860},
  year={2025},
  publisher={Frontiers}
}

@article{tanner2014erps,
  title={ERPs reveal individual differences in morphosyntactic processing},
  author={Tanner, Darren and Van Hell, Janet G},
  journal={Neuropsychologia},
  volume={56},
  pages={289--301},
  year={2014},
  publisher={Elsevier}
}

@inproceedings{surendra2023word,
  title={Word class representations spontaneously emerge in a deep neural network trained on next word prediction},
  author={Surendra, Kishore and Schilling, Achim and Stoewer, Paul and Maier, Andreas and Krauss, Patrick},
  booktitle={2023 International Conference on Machine Learning and Applications (ICMLA)},
  pages={1481--1486},
  year={2023},
  organization={IEEE}
}

@article{bastiaansen2006oscillatory,
  title={Oscillatory neuronal dynamics during language comprehension},
  author={Bastiaansen, Marcel and Hagoort, Peter},
  journal={Progress in brain research},
  volume={159},
  pages={179--196},
  year={2006},
  publisher={Elsevier}
}

@article{schrimpf2020artificial,
  title={Artificial neural networks accurately predict language processing in the brain},
  author={Schrimpf, Martin and Blank, Idan and Tuckute, Greta and Kauf, Carina and Hosseini, Eghbal A and Kanwisher, Nancy and Tenenbaum, Joshua and Fedorenko, Evelina},
  journal={BioRxiv},
  pages={2020--06},
  year={2020},
  publisher={Cold Spring Harbor Laboratory}
}

@article{fedorenko2011functional,
  title={Functional specificity for high-level linguistic processing in the human brain},
  author={Fedorenko, Evelina and Behr, Michael K and Kanwisher, Nancy},
  journal={Proceedings of the National Academy of Sciences},
  volume={108},
  number={39},
  pages={16428--16433},
  year={2011},
  publisher={National Academy of Sciences}
}

@book{pulvermuller2002neuroscience,
  author = {Pulvermüller, Friedemann},
  title = {The neuroscience of language: On brain circuits of words and serial order},
  publisher = {Cambridge University Press},
  year = {2002}
}

@article{goldberg2009nature,
  author = {Goldberg, Adele E.},
  title = {The nature of generalization in language},
  journal = {Cognitive Linguistics},
  volume = {20},
  number = {1},
  pages = {93--127},
  year = {2009}
}

@article{goldberg2003constructions,
  author = {Goldberg, Adele E.},
  title = {Constructions: A new theoretical approach to language},
  journal = {Trends in Cognitive Sciences},
  volume = {7},
  number = {5},
  pages = {219--224},
  year = {2003}
}

@book{goldberg1995constructions,
  author = {Goldberg, Adele E.},
  title = {Constructions: A Construction Grammar Approach to Argument Structure},
  publisher = {University of Chicago Press},
  year = {1995}
}

@book{goldberg2006constructions,
  author = {Goldberg, Adele E.},
  title = {Constructions at Work: The Nature of Generalization in Language},
  publisher = {Oxford University Press},
  year = {2006}
}

@book{goldberg2019explain,
  author = {Goldberg, Adele E.},
  title = {Explain Me This: Creativity, Competition, and the Partial Productivity of Constructions},
  publisher = {Princeton University Press},
  year = {2019}
}

@book{pulvermuller2012meaning,
  author = {Pulvermüller, Friedemann},
  title = {Meaning in the Brain},
  publisher = {MIT Press},
  year = {2012}
}

@article{pulvermuller2021biological,
  author = {Pulvermüller, Friedemann},
  title = {Neural reuse of action perception circuits for language, concepts and communication},
  journal = {Progress in Neurobiology},
  volume = {205},
  pages = {102127},
  year = {2021}
}

@article{henningsen2022modelling,
  author = {Henningsen, K. and Pulvermüller, F.},
  title = {Modelling the processing of argument structure constructions in the brain},
  journal = {Neurobiology of Language},
  volume = {3},
  number = {1},
  pages = {1--23},
  year = {2022}
}

@article{pulvermuller2023neurobiological,
  author = {Pulvermüller, Friedemann},
  title = {Neurobiological mechanisms for generalization and specialization in language},
  journal = {Philosophical Transactions of the Royal Society B},
  volume = {378},
  number = {1875},
  pages = {20210358},
  year = {2023}
}

@article{FIORENTINO201879,
title = {Individual differences in the processing of referential dependencies: Evidence from event-related potentials},
journal = {Neuroscience Letters},
volume = {673},
pages = {79-84},
year = {2018},
issn = {0304-3940},
doi = {https://doi.org/10.1016/j.neulet.2018.02.014},
url = {https://www.sciencedirect.com/science/article/pii/S0304394018300909},
author = {Robert Fiorentino and Lauren Covey and Alison Gabriele}
}

\end{document}